\documentclass[11pt]{article}

\usepackage{latexsym,amsmath,amssymb,amsthm}

\usepackage{graphicx}
\usepackage{epsfig,epstopdf,subcaption}

\usepackage{caption}
\usepackage{tabularx}
\usepackage{color}

\usepackage{comment}

\theoremstyle{definition}

\newcommand\nc{\newcommand}
\newcommand{\bJ}{\mathbf{J}}
\newcommand{\bM}{\mathbf{M}}
\newcommand{\be}{\begin{equation}}
\newcommand{\ee}{\end{equation}}
\newcommand{\ba}{\begin{aligned}}
\newcommand{\ea}{\end{aligned}}
\newcommand{\bea}{\begin{eqnarray}}
\newcommand{\eea}{\end{eqnarray}}
\newcommand{\bE}{{\mathbf{E}}}
\newcommand{\bH}{{\mathbf{H}}}

\nc\bn{{\mathbf{n}}}
\nc\bphi{{\boldsymbol{\phi}}}
\nc\bpsi{{\boldsymbol{\psi}}}

\nc\br{{\mathbf{r}}}
\nc\curl{\nabla\times}
\nc\dive{\nabla\cdot}
\nc\pt{\frac{\partial}{\partial t}}
\nc\pet{\frac{\partial \bE}{\partial t}}
\nc\pht{\frac{\partial \bH}{\partial t}}

\begin{document}
\title{A Quantitative Study of the Effect of Cladding Thickness
  on Modal Confinement Loss in Photonic Waveguides}
  
\author{ Shidong Jiang
\thanks{Department of Mathematical Sciences, New Jersey Institute of
Technology, Newark, New Jersey 07102
({\tt shidong.jiang@njit.edu}).}
\and
Jun Lai
\thanks{Courant Institute of Mathematical Sciences, New York University, New York, New York, 10012
({\tt lai@cims.nyu.edu}).}
}



\maketitle
\begin{abstract}
There has been increasing interest in making the photonic devices more and
more compact in the integrated photonics industry, and
one of the important questions for manufacturers and design engineers is
how to quantify the effect of the finite cladding thickness on the
modal confinement loss of photonic waveguides. This requires
at least six to seven digits accuracy for the computation of propagation constant
$\beta$
since the modal confinement loss is proportional to the imaginary part of $\beta$
that is six to seven orders of magnitude smaller than its real part by the industrial standard.
In this paper, we present an accurate and efficient method to compute 
the propagation constant of electromagnetic modes of photonic waveguides 
with arbitrary number of (nonsmooth) inclusions in a layered media. 
The method combines 
a well-conditioned boundary integral equation formulation for 
photonic waveguides which requires the discretization of the material interface
only, and efficient Sommerfeld integral representations to 
treat the effect of the layered medium. Our scheme is capable of calculating 
the propagation loss of the electromagnetic modes with high fidelity, even
for waveguides with corners imbedded in a cladding material of finite thickness.
The numerical results, with more than $10$-digit accuracy,
show quantitatively that the modal confinement loss of the rectangular waveguide
increases exponentially fast as the cladding thickness decreases.
\end{abstract}

\section{Introduction}

In many photonic devices, the input and output channels take the form of
(approximately) straight waveguides with uniform cross sections. It is well known
that such straight waveguide structure support only a finite number of 
electromagnetic modes which can propagate with little or no energy loss
for a given frequency. This digitizes the design of photonic devices and reduces a
seemingly infinite dimensional problem to a finite dimensional one, which greatly
simplifies the design process. The problem of mode calculation is concerned
with characterizing the nature of 
the propagating waves for a waveguide of given cross-section, including their 
propagation constant and the structure of the associated electromagnetic fields.

Recently, as the integrated optical industry has grown, engineers
have been trying to assemble more and more photonic components into a
single chip. As a result, the effect of finite cladding thickness on the
performance of the individual photonic components requires more care
and more accurate modeling.
For photonic waveguides, some obvious but important question are as
follows: how does the thickness of the cladding alter/affect the propagation
constant of each mode? What is the minimal cladding thickness for a given
propagation loss threshold? These problems are naturally cast as nonlinear
eigenvalue problems in a layered medium and present new challenges for numerical
simulations, as they involves boundary conditions along the infinitely long
material interface separating different layers.

Popular methods for mode calculation such as finite difference \cite{fd2,fd3}
or finite element methods \cite{fem5,fem6} 
require the discretization of a finite domain,
supplemented by artificial boundary conditions, such as perfectly matched
layers, to simulate the effect of an infnite medium.
Using these methods to study the effect of cladding thickness 
on propagation loss is problematic, since the computational domain has to be much
larger than the waveguide cross-section in order to take the effect of layers into 
account. Existing boundary integral
methods \cite{srep,leslie1,ferrando,lu2,pone} are capable of calculating the propagation
constant to high accuracy when the boundary of the waveguide consists of 
smooth curves in the absence of layers, but they become ill-conditioned
and inaccurate for 
waveguides with nonsmooth geometry such as the rectangular waveguides considered
in this paper and less efficient when the
effect of layers has to be included. At the same time, the modal confinement
loss is connected with the imaginary part of the propagation constant of the
electromagnetic modes, which is very often at least six to seven orders smaller
than the real part of the propagation constant. Thus, one needs at least six to
seven digits accuracy in the overall computation in order to obtain a single
significant digit for the modal confinement loss. This high accuracy demand
presents a great challenge, requiring that the scheme be
well-conditioned, high-order, and efficient, so as not to consume excessively
large amounts of computational resources and for it to be of practical use in 
design.

In this paper, we present an integral formulation for the electromagnetic mode
calculation of photonic waveguides in layered media. The formulation
combines a carefully chosend  boundary integral representation \cite{skie} 
for photonic waveguides and an efficient Sommerfeld 
integral representation \cite{lai} for layers.
The overall numerical scheme is robust, high-order, and efficient. We demonstrate
the performance of the scheme by showing that the effect of finite cladding on 
modal confinement loss of rectangular dielectric waveguides increases 
exponentially fast as the thickness decreases. 

\section{Integral Formulation for the Mode Calculation}
\label{sec:ieformulation}
\subsection{Notation}
\begin{figure}[htbp]
\centering
\includegraphics[width=0.7\linewidth]{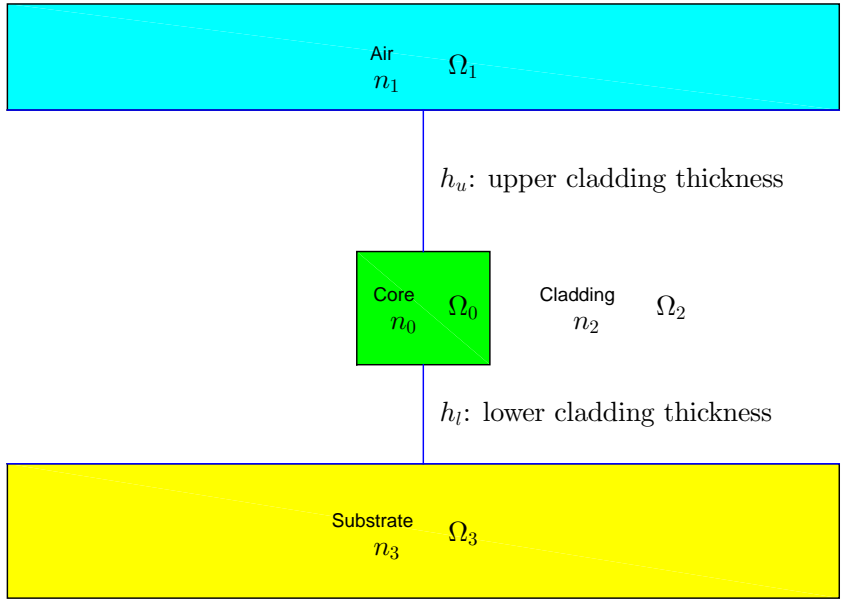}
\caption{Cross section of a rectangular waveguide in layered medium.}
\label{fig:waveguide}
\end{figure}

We assume that electromagnetic fields are propagated along the $z$-axis, and that
the geometric structure of the photonic waveguides in a layered medium is
completely determined by its cross section in the $xy$-plane (or $\mathbb{R}^2$)
shown in Fig. \ref{fig:waveguide}.
We denote the top layer (air for dielectric waveguides) by $\Omega_1$ and its
index of refraction by $n_1$, the cladding domain by $\Omega_2\in\mathbb{R}^2$
and its index of refraction by $n_2$, and the bottom substrate domain
by $\Omega_3\in\mathbb{R}^2$ and its index of refraction by $n_3$, 
respectively. The cross section of the rectangular waveguide is denoted by
$\Omega_0$ with $n_0$ the index of refraction of the core. The boundary of the
waveguide is denoted by $\Gamma_0$ with $\nu$ the unit outward normal vector
and $\tau$ the unit tangential vector, respectively. Two horizontal lines at
$y=y_t=0$
and $y=y_b=-h_u-h-h_l$ separating the top and bottom layers from the cladding
are denoted by $\Gamma_t$ and $\Gamma_b$, respectively. Here $h$ is the height of
the rectangular waveguide, $h_u$ and $h_l$ are the upper and lower cladding
thickness, respectively.
\subsection{Original PDE Formulation}
The electromagnetic field satisfies the Maxwell equations in each domain:
\be\label{maxwell}
\left\{\ba \curl \bE&=-\mu_0 \pht, & \dive \bE=0, \\
\curl \bH&=\epsilon_0 n^2 \pet, & \dive \bH=0,\ea\right.
\ee
where $n$ is the index of refraction of the domain.
In the mode calculation, the fundamental assumption is that the electromagnetic
field takes the following form:
\be\label{modeassump}
[\bE(x,y,z,t),\ \bH(x,y,z,t)] =[\bE(x,y),\ \bH(x,y)] e^{i(\beta z-\omega t)},
\ee
where $\beta$ is the propagation constant and $\omega$ is the frequency of the
incident wave.
Combining this assumption with the Maxwell equations, we observe that every
component of $\bE(x,y)$ and $\bH(x,y)$ satisfies 
the two dimensional Helmholtz equation in each domain:
\be\label{helmholtzeq}
\left[\Delta + (k^2-\beta^2)\right] u = 0, 
\ee
where $k=nk_v$, and $k_v=\omega\sqrt{\epsilon_0\mu_0}=\omega/c$ is the wave number
in vacuum. On the material interface, the boundary conditions are that the
tangential components of the electromagnetic fields are continuous.
This leads to the following four boundary conditions on each boundary:
\be\label{modebc}
[E_z] = 0, \quad [H_z] = 0, \quad
[E_\tau]=0, \quad [H_\tau]=0,
\ee
where $[\cdot]$ denotes the jump of the quantity across the boundary.

\subsection{Layer Potentials, Sommerfeld Integral and Representation}
Let $\Omega$ be a bounded domain in $\mathbb{R}^2$ with boundary $\Gamma$.
We denote the points in $\mathbb{R}^2$ by $P$ and $Q$. Here $Q$ is usually a
point (the source point) on the boundary $\Gamma$, and
$P$ is in general an arbitrary point (the target point) in $\mathbb{R}^2$.
The Green's function for \eqref{helmholtzeq} is
\be\label{greenfun}
G^k(P,Q)=\frac{i}{4}H_0^{(1)}\left(\sqrt{k^2-\beta^2}\|P-Q\|\right),
\ee
where $H_0^{(1)}$ is the Hankel function of the first kind of order zero
\cite{olver_nist_2010}. Let $\sigma$ be a square integrable function on $\Gamma$.
We define the single, double, and anti-double
layer potentials by the following formulas, respectively:
\bea\label{layerpot}
\left\{\ba S_\Gamma^k[\sigma](P)&=\int_\Gamma G^k(P,Q)\sigma(Q)ds_Q,\\
D_\Gamma^k[\sigma](P)&=\int_\Gamma \frac{\partial G^k(P,Q)}{\partial \nu(Q)}
\sigma(Q)ds_Q,\\
T_\Gamma^k[\sigma](P)&=\int_{\Gamma}
\frac{\partial G^k(P,Q)}{\partial \tau(Q)}\sigma(Q)ds_Q.
\ea\right.
\eea
Let $P = (x,y)$ and $Q=(x_0,y_0)$. Then the Green's function in \eqref{greenfun}
has the following Sommerfeld integral representation \cite{lai}:
\begin{equation}\label{sommer}
  G^k(P,Q) = \frac{1}{4\pi}\int_{-\infty}^{\infty}
  \frac{e^{-\sqrt{\lambda^2-{k}^2+\beta^2}|y-y_0|}}
  {\sqrt{\lambda^2-{k}^2+\beta^2}} e^{i\lambda(x-x_0)}d\lambda.
\end{equation}
Using the above representation, the layer potentials defined in
\eqref{layerpot} and their derivatives have an alternative representation for sources lying on the interface of a layered medium. For example, the single, double, and anti-double layer potentials for the top layer
with sources on $\Gamma_t$ ($y=y_t$) have the following Sommerfeld representations:
\be\label{singlesommer}
\left\{\ba \widehat{S}_{\Gamma_t}^{k_1}[\widehat{\sigma}](P)&= \frac{1}{4\pi}\int_{-\infty}^{\infty}
\frac{e^{-\sqrt{\lambda^2-{k_1}^2+\beta^2}|y-y_t|}}
{\sqrt{\lambda^2-{k_1}^2+\beta^2}} e^{i\lambda x}\widehat{\sigma}(\lambda)d\lambda,\\
\widehat{D}_{\Gamma_t}^{k_1}[\widehat{\sigma}](P) & = \frac{1}{4\pi}\int_{-\infty}^{\infty}
e^{-\sqrt{\lambda^2-{k_1}^2+\beta^2}|y-y_t|}
 e^{i\lambda x}\widehat{\sigma}(\lambda)d\lambda, \\
\widehat{T}_{\Gamma_t}^{k_1}[\widehat{\sigma}](P) & = \frac{1}{4\pi}\int_{-\infty}^{\infty}
\frac{i\lambda e^{-\sqrt{\lambda^2-{k_1}^2+\beta^2}|y-y_t|}}
{\sqrt{\lambda^2-{k_1}^2+\beta^2}} e^{i\lambda x}\widehat{\sigma}(\lambda)d\lambda.
\ea\right.
\ee
where $\widehat{\sigma}$ is the Fourier transform of the unknown density $\sigma$ on $\Gamma_t$.
The advantage of the Sommerfeld representation is that the kernels in
\eqref{singlesommer} decay exponentially fast as $\lambda$ approaches infinity
(while the kernels using the Green's function in \eqref{layerpot} decays only algebraically).

\subsection{Integral Representations of the Electromagnetic Fields}
Suppose that $\bJ(x,y)=\bJ_z\widehat{k}+\bJ_\tau \tau$ and
$\bM(x,y)=\bM_z\widehat{k}+\bM_\tau \tau$
are two unknown surface currents on $\Gamma$.
Following \cite{skie}, we define two vector fields $\bphi_\Gamma^k[\bJ,\bM]$,
$\bpsi_\Gamma^k[\bJ,\bM]$ by the following formulas:
\be\label{mullere}
\left\{\ba
\bphi_{\Gamma,x}^k[\bJ,\bM]
&= -\frac{1}{ik_v}
\frac{\partial}{\partial x}T_\Gamma^k[J_\tau]
+\frac{\beta}{k_v}
\frac{\partial}{\partial x}S_\Gamma^k[J_z]
+\frac{k^2}{ik_v}S_\Gamma^k[J_\tau\tau_x]\\
&\ -\frac{\partial}{\partial y}S_\Gamma^k[M_z]+i\beta S_\Gamma^k[M_\tau \tau_y],\\
\bphi_{\Gamma,y}^k[\bJ,\bM] &= -\frac{1}{ik_v}
\frac{\partial}{\partial y}T_\Gamma^k[J_\tau]
+\frac{\beta}{k_v} \frac{\partial}{\partial y}S_\Gamma^k[J_z]
+\frac{k^2}{ik_v}S_\Gamma^k[J_\tau\tau_y]\\
&\ +\frac{\partial}{\partial x}S_\Gamma^k[M_z]-i\beta S_\Gamma^k[M_\tau \tau_x],\\
\bphi_{\Gamma,z}^k[\bJ,\bM] &= -\frac{\beta}{k_v} T_\Gamma^k[J_\tau]
+\frac{(k^2-\beta^2)}{ik_v}
S_\Gamma^k[J_z]+D_\Gamma^k[M_\tau],
\ea\right.\ee
and
\be\label{mullerh}
\left\{\ba
\bpsi_{\Gamma,x}^k[\bJ,\bM] &= \frac{1}{ik_v}\frac{\partial}{\partial x}
T_\Gamma^k[M_\tau]
-\frac{\beta}{k_v} \frac{\partial}{\partial x}S_\Gamma^k[M_z]
-\frac{k^2}{ik_v}S_\Gamma^k[M_\tau\tau_x]\\
&\ -\frac{k^2}{k_v^2}\frac{\partial}{\partial y}S_\Gamma^k[J_z]
+i\beta \frac{k^2}{k_v^2} S_\Gamma^k[J_\tau \tau_y],\\
\bpsi_{\Gamma,y}^k[\bJ,\bM] &= \frac{1}{ik_v}\frac{\partial}{\partial y}
T_\Gamma^k[M_\tau]
-\frac{\beta}{k_v} \frac{\partial}{\partial y}S_\Gamma^k[M_z]
-\frac{k^2}{ik_v}S_\Gamma^k[M_\tau\tau_y]
\\&\ +\frac{k^2}{k_v^2}\frac{\partial}{\partial x}S_\Gamma^k[J_z]
-i\beta \frac{k^2}{k_v^2} S_\Gamma^k[J_\tau \tau_x],\\
\bpsi_{\Gamma,z}^k[\bJ,\bM] &= \frac{\beta}{k_v} T_\Gamma^k[M_\tau]
-\frac{(k^2-\beta^2)}{ik_v} S_\Gamma^k[M_z]+\frac{k^2}{k_v^2}D_\Gamma^k[J_\tau].
\ea\right.\ee
When $\Gamma$ is a horizontal line, we also define similar expressions for
$\widehat{\bphi}_{\Gamma}^k[\widehat{\bJ},\widehat{\bM}]$,
$\widehat{\bpsi}_{\Gamma}^k[\widehat{\bJ},\widehat{\bM}]$ with the associate layer
potentials replaced by their corresponding Sommerfeld representations.
We now propose the following representations for the electromagnetic fields in
various regions for the mode calculation of photonic waveguide in a layered
medium shown in Fig.~\ref{fig:waveguide}.

\be\label{representation}
\left\{
\ba
\left(\bE,\, \bH\right) &= \left(\bphi_{\Gamma_0}^{k_0}[\bJ,\bM],\,
\bpsi_{\Gamma_0}^{k_0}[\bJ,\bM]\right) \quad \text{in}\,\, \Omega_0,\\
\left(\bE,\, \bH\right) &= \left(\widehat{\bphi}_{\Gamma_t}^{k_1}
     [\widehat{\bJ}_t,\widehat{\bM}_t],\,
     \widehat{\bpsi}_{\Gamma_t}^{k_1}[\widehat{\bJ}_t,\widehat{\bM}_t]\right)
     \quad \text{in}\,\, \Omega_1,\\
     \left(\bE,\, \bH\right) &= \left(\widehat{\bphi}_{\Gamma_t}^{k_2}
          [\widehat{\bJ}_t,\widehat{\bM}_t],\,
\widehat{\bpsi}_{\Gamma_t}^{k_2}[\widehat{\bJ}_t,\widehat{\bM}_t]\right)\\
&+\left(\bphi_{\Gamma_0}^{k_2}[\bJ,\bM],\,
\bpsi_{\Gamma_0}^{k_2}[\bJ,\bM]\right)\\
&+\left(\widehat{\bphi}_{\Gamma_b}^{k_2}[\widehat{\bJ}_b,\widehat{\bM}_b],\,
\widehat{\bpsi}_{\Gamma_b}^{k_2}[\widehat{\bJ}_b,\widehat{\bM}_b]\right)
\quad \text{in}\,\, \Omega_2,\\
\left(\bE,\, \bH\right) &= \left(\widehat{\bphi}_{\Gamma_b}^{k_3}
     [\widehat{\bJ}_b,\widehat{\bM}_b],\,
     \widehat{\bpsi}_{\Gamma_b}^{k_3}[\widehat{\bJ}_b,\widehat{\bM}_b]\right)
     \quad \text{in}\,\, \Omega_3.
\ea
\right.
\ee

In other words, the fields inside the core are generated by the unknown densities
$\bJ$ and $\bM$ on its boundary $\Gamma_0$ via the representations
\eqref{mullere}-\eqref{mullerh}. The fields in the top layer are generated by the
unknown densities $\widehat{\bJ}_t$ and $\widehat{\bM}_t$ in the Fourier domain
via the Sommerfeld representation of \eqref{mullere}-\eqref{mullerh}. The
fields in the bottom layer are generated by the unknown densities
$\widehat{\bJ}_b$ and
$\widehat{\bM}_b$ in the Fourier domain. Finally, the fields in the cladding region
are generated by all six unknown vector densities via suitable representations.

It is tedious, yet straightforward to show \cite{skie}
that the above representation satisfies the
corresponding Helmholtz equation and also the Maxwell equations in each region.
The boundary conditions on $\Gamma_t$, $\Gamma_0$, and $\Gamma_b$
together with the well-known jump relations of the layer potentials (see, for
example, \cite{kress,skie}) then lead to a $12\times 12$ block system $Ax=0$,
where  $x=[\widehat{J}_{t,x}\, \widehat{J}_{t,z}\, \widehat{M}_{t,x}\,
  \widehat{M}_{t,z}\,J_\tau\, J_z\,
  M_\tau, M_z,\widehat{J}_{b,x}\, \widehat{J}_{b,z}\, \widehat{M}_{b,x}\,
  \widehat{M}_{b,z}]^T$.
Similar arguments in \cite{skie} show that $A$ is a second kind Fredholm integral
operator for smooth boundaries. And the propagation constant $\beta$ of the eletromagnetic mode is a
complex number for which the integral operator $A$ has a nontrivial nullspace.
We would like to emphasize again that our formulation is well-conditioned and thuse
capable of achieving high accuracy even in the case of nonsmooth geometries,
say, waveguides with corners, while it is difficult to obtain high accuracy
for nonsmooth cases using existing boundary integral methods
\cite{srep,leslie1,ferrando,lu2,pone}
due to the intrinsic ill-conditioning of their formulations.

\section{Numerical Algorithm}
\subsection{Discretization of the Layer Potentials}
The layer potentials involve weakly singular integrals and many photonic waveguides
in integrated optics are of rectangular shape. Thus we need to deal with the
singularities in the kernel and the densities (induced by the corner singularity
in the geometry). 
Here we use a collocation
Nystr{\"{o}}m method with dyadic refinement toward the corners to discretize the
layer potentials. 

We first divide each side of the rectangle into $N_m+2$ subintervals of equal
length, then divide each end subinterval dyadically into $N_e$ smaller and
smaller subintervals. On each subinterval, we place $p$ shifted and scaled
Gauss-Legendre nodes and the solution is approximated by a polynomial
of degree less than $p$. So the total number of discretization points $N$ on
each side is $p\cdot(N_m+2N_e)$.
For each collocation point, the integrals in the layer
potentials are discretized via either a precomputed generalized Gaussian quadrature
when they are weakly or nearly singular or regular Gaussian quadrature when
they are smooth.
\subsection{Discretization of the Sommerfeld Representation}
\begin{figure}[htbp]
\center
\includegraphics[scale=0.5]{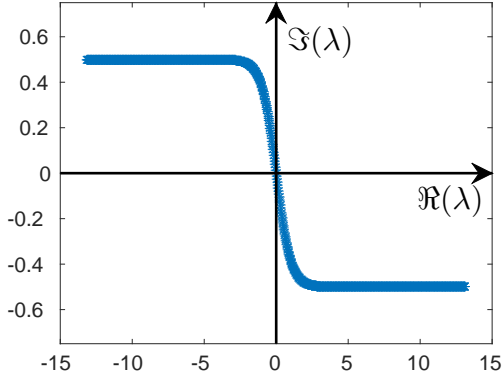}
\caption{The hyperbolic tangent contour used in the Sommerfeld representation}
\label{fig:contour}
\end{figure}

To numerically evaluate the Sommerfeld integral to high order, we need to avoid
the square root singularity in \eqref{sommer}.  This can be achieved by contour
deformation. In particular, we choose the following hyperbolic tangent contour
(see Figure \ref{fig:contour})
\begin{equation}
\lambda = t-\frac{\tanh(t)}{2}i, \quad t\in \mathbb{R}.
\end{equation}
We then apply a truncated trapezoidal rule to discretize the Sommerfeld integrals,
which achieves spectral accuracy due to the smoothness and exponential decay of the
integrand. Specifically, we truncate $t$ at $[-T,T]$ and discretize $t$
uniformly by $N_S$ points on that interval. The value of $T$ depends on the wave
numbers and the distance between the core of the waveguide and the interface
separating the layers. In our numerical experiments, we set
$T=13$, after which the contribution from Sommerfeld integral is exponentially small.
We refer the reader to \cite{lai} for a more
detailed discussions of the advantages of the Sommerfeld representation as compared
with the original representation using Green's function directly.
\subsection{Finding Propagation Constant via Root Finding}
We apply the method in \cite{leslie1} to find the propagation constant
(or the effective index $n_e=\beta/k_v$ in the actual implementation).
Suppose that $M(\beta)$ is the resulting matrix from the discretization.
The propagation constant is obtained by finding the roots of a scalar function
$f(\beta)=1/\left(u^TM^{-1}(\beta) v\right)$ via M\"{u}ller's method \cite{dmuller}.
Here $u$ and $v$ are two fixed random column vectors.
\section{Numerical Results}
In this section, we present some benchmark calculation on the effect of
lower cladding thickness on the modal confinement loss of rectangular
dielectric waveguides.

{\bf Example 1: a high refractive index contrast silica waveguide.}
In this example, the cross section of the
waveguide is of square shape with the side length equal to $3.4\mu m$.
The refractive index of the cladding is $n_0=1.4447$, while that of the core
is $2\%$ higher, i.e., $n_2=1.4447\times 1.02$. The refractive indices of the
silicon base and the air are $3.476$, and $1.0003$, respectively. 
The wavelength of the incident field is $1550nm$. In the simplified model where
the top and bottom layers are absent, our calculation in \cite{skie} shows that
the waveguide supports a mode with double degeneracy with the effective index
$n_e\simeq 1.458601414886$. The result is accurate to $12$ digits (see \cite{skie}
for details).

We first check the convergence rate of our numerical scheme.
Table \ref{tab_convergence} shows the effective index of the first mode
found by our scheme for
various number of discretization points on each side of the square when
$h_l=4\mu m$. The number of points in the Sommerfeld representation
is fixed $N_S=1000$ as we found increasing $N_S$ will give about the same
values under double precision computation. We observe that $12$ digit accuracy is
achieved with $N=500$ (to be more precise,
$N_m = 10$, $N_e = 20$, $p = 10$)
points. Thus we set $N=500$ for our subsequent calculation.

\begin{table}[htbp]
\begin{center}
    \resizebox{0.5\linewidth}{!}{%
\begin{tabular}{|c|c|c|} \hline
N & Real & Imaginary  \\\hline
200 & 1.45860122763585 & 2.673537E-8 \\\hline
300 & 1.45860122756971 & 2.498713E-8 \\\hline
400 & 1.45860122756758 & 2.493100E-8 \\\hline
500 & 1.45860122756751 & 2.492917E-8 \\\hline
600 & 1.45860122756751 & 2.492915E-8 \\\hline
\end{tabular}}
\end{center}
\caption{Convergence study for the effective index of the second mode
  when the lower cladding thickness is $4 \mu m$. The first column lists
  the number of discretization points on each side of the square; the second
  and third columns list the real and imaginary parts of the effective index of
  the second mode.}
\label{tab_convergence}
\end{table}

We now consider the effect of the finite thickness of the cladding. To simplify
our discussion, the upper thickness of the cladding is fixed at $15\mu m$, while
the thickness of the lower cladding $h_l$ is varied from $10\mu m$ to $4\mu m$. The
presence of the layers destroys the symmetry of the square waveguide, and the
doubly degenerate mode is split into two single modes.

Table~\ref{tab:NumericalResult1} lists the effective indices of two modes for
various lower cladding thickness. The results are obtained by setting  $N_S=1000$ for Sommerfeld integrals and $N=500$ 
for each side of the 
waveguide, which achieves about $12$-digit accuracy by the aforementioned
convergence study.
\begin{table}[htbp]
  \begin{center}
    \resizebox{0.8\linewidth}{!}{%
\begin{tabular}{|c|c|c|c|c|} \hline
  &  \multicolumn{2}{|c|}{First mode} & \multicolumn{2}{|c|}{Second mode}  \\\hline
$h_l$ & $Re(n_e)$ & $Im(n_e)$ & $Re(n_e)$ & $Im(n_e)$  \\\hline
4  & 1.45860122757 & 2.4929E-8 &
1.45860127384 & 1.26765E-7 
\\\hline
5 & 1.45860138033 & 4.578E-9 &
 1.45860138880 & 2.3309E-8 
\\\hline
6 & 1.45860140847 & 8.47E-10 &
 1.45860141003 & 4.317E-9 
\\\hline
7  & 1.45860141369 & 1.57E-10&
1.45860141398 & 8.04E-10 
\\\hline
8  & 1.45860141466 & 2.9E-11&
1.45860141471 & 1.50E-10 
\\\hline
9  & 1.45860141484 & 5E-12&
1.45860141485 & 2.8E-11  
\\\hline
10 & 1.45860141488 & 1E-12&
1.45860141488 & 5E-12   
\\\hline
\end{tabular}}
\end{center}
\caption{Effective indices versus lower cladding thickness $h_l$ for Example 1.}
\label{tab:NumericalResult1}
\end{table}

\begin{figure}[htbp]
\center
\includegraphics[scale=0.65]{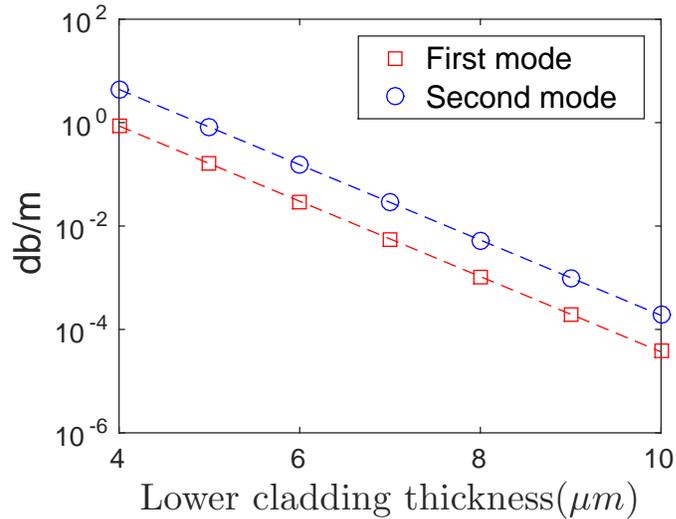}
\caption{Modal confinement loss versus the lower cladding thickness for Example 1.
  The $y$-axis
uses logarithmic scale (base $10$). Dashed lines are the least squares fit to the data.}
\label{fig:NumericalResult1}
\end{figure}

We now calculate the modal confinement loss (in dB/m) \cite{tpwhite1}
via the formula
\be
L=\frac{20}{\ln(10)}\cdot \frac{2\pi}{\lambda}\cdot Im(n_e)\cdot 10^9
\ee
with $\lambda$ the vacuum wavelength measured in nanometers.
Figure \ref{fig:NumericalResult1} plots out
the modal confinement loss $L$ versus $h_l$ and their least squares fits.

The least squares fit of both data sets shows that the slope is about
$K_1\approx -0.728 \mu m^{-1}$. That is, the modal confinement loss $L$ increases
exponentially fast as $h_l$ is decreasing:
\be
L \propto 10^{K_1 h_l}.
\ee

This can be explained as follows. The electromagnetic fields
decays in the cladding region in the form ${\bf F} \propto
e^{-2\pi\sqrt{n_2^2-n_e^2}|y|/\lambda}$.
Thus, the energy loss, which is directly linked to the modal confinement loss,
due to the finite lower cladding thickness is proportional
to $|{\bf F}|^2\propto e^{-4\pi\sqrt{n_2^2-n_e^2} h_l/\lambda}=10^{-4\pi \sqrt{n_2^2-n_e^2}
  (\log_{10}e) h_l/\lambda}$. With $n_2=1.4447\times 1.02$, $n_e\approx1.4586$, $\lambda=1.55\mu m$, we have
\be
S_1 = -4\pi \sqrt{n_2^2-n_e^2}\cdot (\log_{10}e)/ \lambda \approx -0.738 \mu m^{-1},
\ee
which is in good agreement with the value of $K_1$ (about $1.3\%$ relative error). 
%
%

\begin{figure}[htbp]
  \captionsetup[subfigure]{justification=centering}
  \centering
\begin{subfigure}{.45\textwidth}
  \centering
  \includegraphics[width=0.99\linewidth]{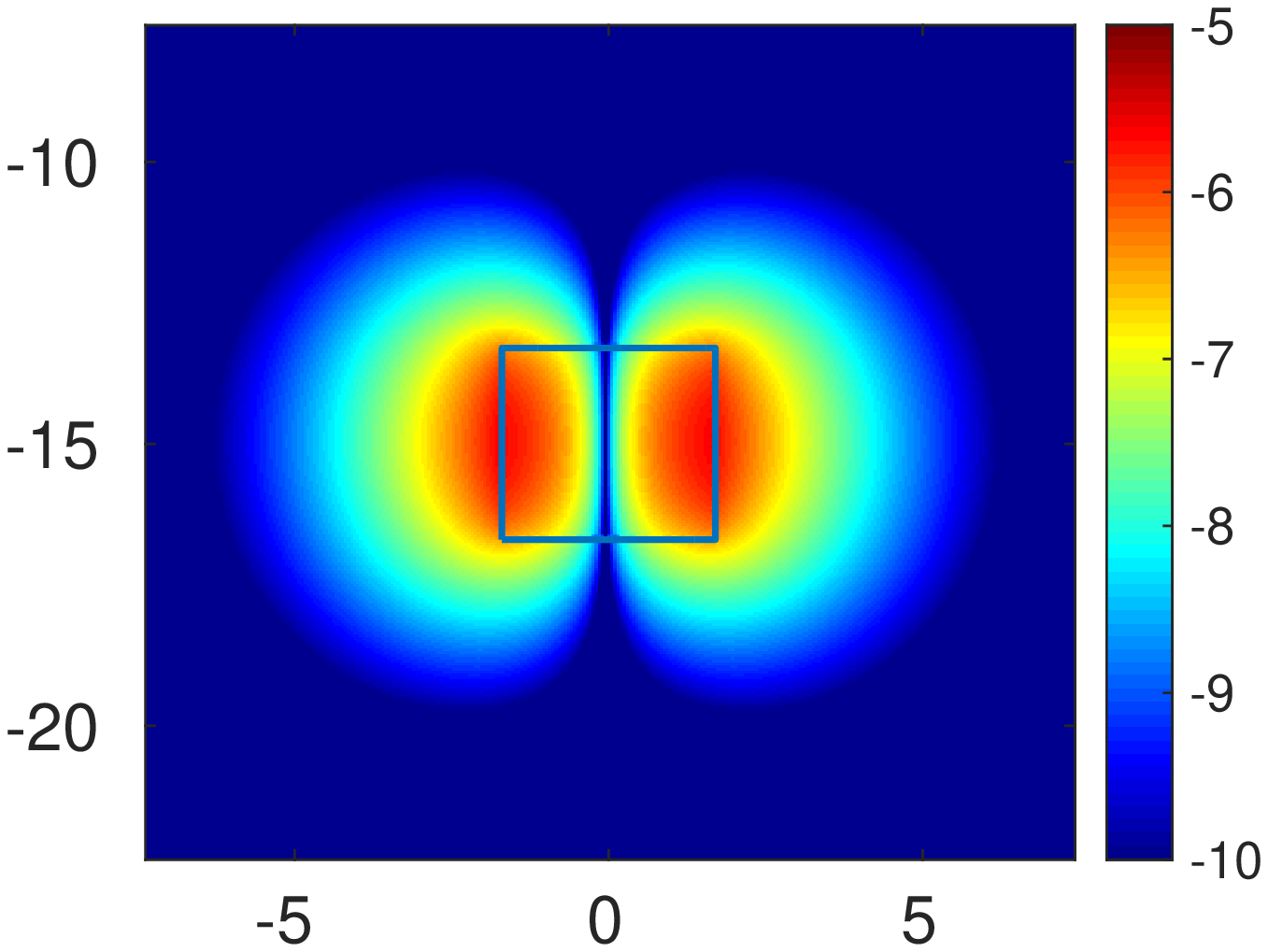}
  \centering
  \caption{$|E_z(x,y)|$ of the first mode}
\end{subfigure}%
\begin{subfigure}{.45\textwidth}
  \centering
  \includegraphics[width=0.99\linewidth]{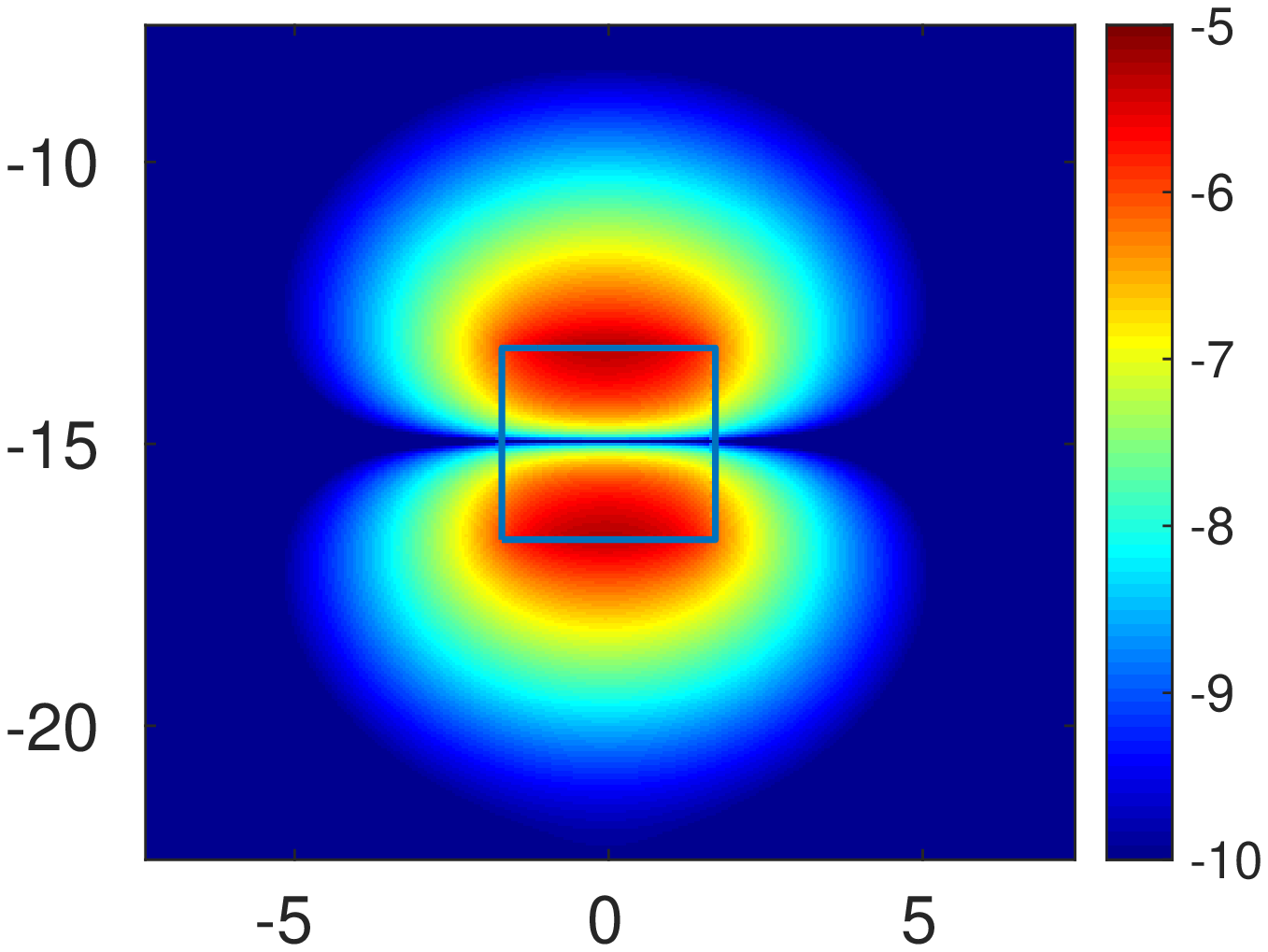}
  \centering
  \caption{$|H_z(x,y)|$ of the first mode}
\end{subfigure}
\begin{subfigure}{.45\textwidth}
  \centering
  \includegraphics[width=0.99\linewidth]{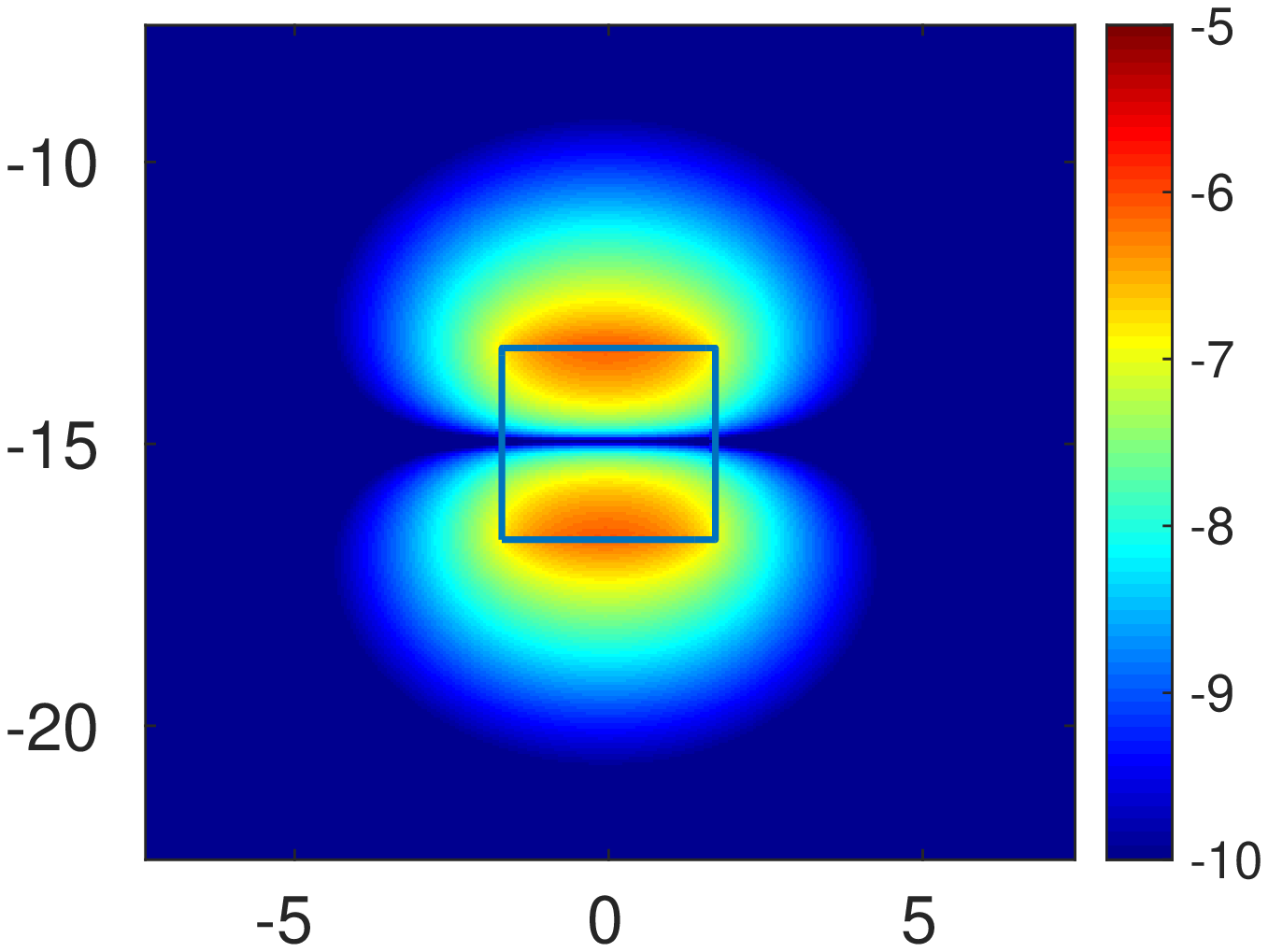}
  \centering
  \caption{$|E_z(x,y)|$ of the second mode}
\end{subfigure}%
\begin{subfigure}{.45\textwidth}
  \centering
  \includegraphics[width=0.99\linewidth]{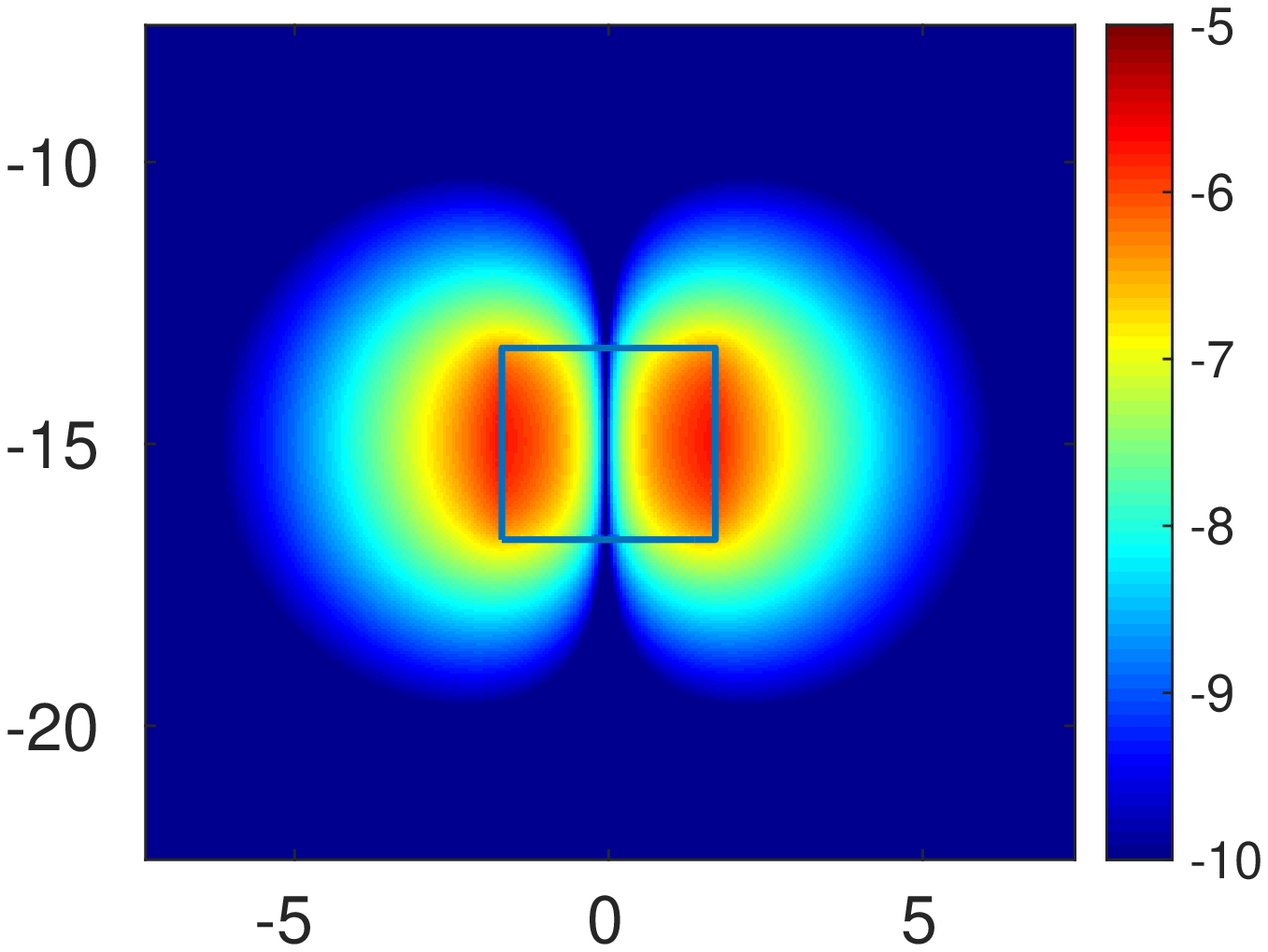}
  \centering
  \caption{$|H_z(x,y)|$ of the second mode}
\end{subfigure}  
\caption{Magnitude of the electromagnetic field $E_z$ and $H_z$ of the first and second modes with $h_l = 4\mu m$ in
  Example 1. The square represents the boundary of the waveguide. The colorbar uses logarithmic scale.}
\vspace{-3.0mm}
\label{fig:NumericalResult2}
\end{figure}

%
We have plotted the magnitude of the electromagnetic field $|E_z|$ and $|H_Z|$
of the two modes in Figure \ref{fig:NumericalResult2}. We observe that the field is
concentrated near the core and decays exponentially fast away from the core,
which confirms our previous claim about the behavior of the field in the cladding
region.

{\bf Example 2: a low refractive index contrast silica waveguide.} In this example,
the cross section of the waveguide
is of square shape with the side length equal to $5.2\mu m$. The refractive indices of the cladding, core,
silicon base, and the air are $1.4447$, $1.4447\times 1.0075$, $3.476$, and $1.0003$, respectively. The wave length
of the incident field is $1550nm$. The upper thickness of the cladding is fixed at $15\mu m$. And we vary
the thickness of the lower cladding from $4\mu m$ to $12\mu m$. The discretization is the same as Example 1. Table~\ref{tab:NumericalResult2} lists the effective indices of two modes for
various lower cladding thickness. 
\begin{table}[htbp]
\begin{center}
\resizebox{0.8\linewidth}{!}{%
\begin{tabular}{|c|c|c|c|c|} \hline
  &  \multicolumn{2}{|c|}{First mode} & \multicolumn{2}{|c|}{Second mode}  \\\hline
$h_l$ & $Re(n_e)$ & $Im(n_e)$ & $Re(n_e)$ & $Im(n_e)$  \\\hline
4  & 1.449463952659 & 1.34678E-7&
1.449464113248 & 7.43394E-7 
\\\hline
5 &1.449465037025 & 4.9161E-8&
  1.449465094948 & 2.71536E-7 
\\\hline
6 &1.449465433894 & 1.8051E-8&
  1.449465454711 & 9.9754E-8 
\\\hline
7  & 1.449465579918 & 6.659E-9&
1.449465587215 & 3.6813E-8 
\\\hline
8  & 1.449465633876 & 2.465E-9&
1.449465636215 & 1.3635E-8
\\\hline
9  & 1.449465653883 & 9.15E-10&
1.449465654396 & 5.066E-9 
\\\hline
10 & 1.449465661324 & 3.41E-10&
1.449465661161 & 1.887E-9
\\\hline
11  &1.449465664098 & 1.27E-10&
1.449465663685 & 7.04E-10 
\\\hline
12  &1.449465665135 & 4.7E-11&
1.449465664628 & 2.63E-10 
\\\hline
\end{tabular}}
\end{center}
\caption{Effective indices of the photonic waveguide in Example 2.}
\label{tab:NumericalResult2}
\end{table}

\begin{figure}
\center
\includegraphics[scale=0.65]{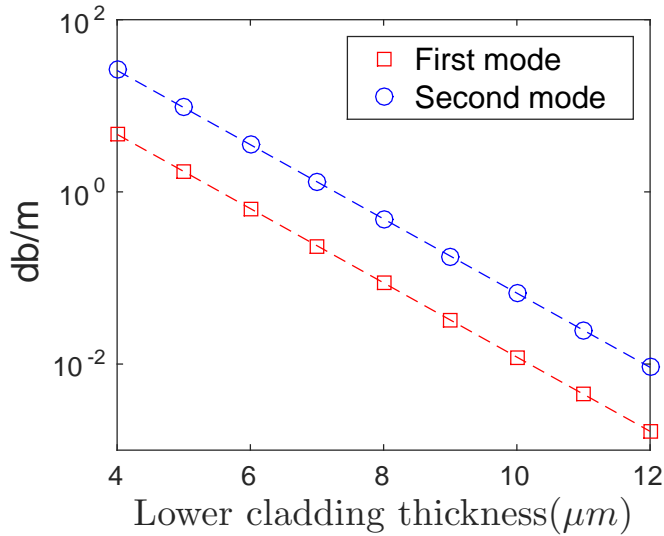}
\caption{Modal confinement loss versus the lower cladding thickness for Example 2.
  The $y$-axis
uses logarithmic scale (base $10$). Dashed lines are the least squares fit to the data.}
\label{fig:NumericalResult4}
\end{figure}

Figure \ref{fig:NumericalResult4} shows the propagation loss $L$ with
respect to the lower cladding thickness and their least squares line fit.

The least squares line fit of both data sets shows that the slope is about
$K_2\approx -0.431 \mu m^{-1}$. That is, the modal confinement loss $L$ increases
exponentially fast as $h_l$ is decreasing:
\be
L \propto 10^{K_2 h_l}.
\ee
Similar argument as in Example 1 shows that 
 the energy loss
due to the finite lower cladding thickness is proportional
to $|{\bf F}|^2\propto 10^{S_2 h_l}$. With $n_2=1.4447\times 1.0075$,
$n_e\approx1.449465$, $\lambda=1.55\mu m$, we have
\be
S_2 = -4\pi \sqrt{n_2^2-n_e^2}\cdot (\log_{10}e)/ \lambda \approx -0.468 \mu m^{-1},
\ee
which is in good agreement with the value of $K_2$ (about $7.8\%$ relative error). 
A comparison of these two examples show that high contrast silica waveguides not only
have smaller core size, but also need much thinner cladding for the same modal
confinement loss. Hence, more compact photonic devices can be made using high contrast
silica waveguides.
\section{Conclusions}
We have presented a well-conditioned integral formulation for
the calculation of
electromagnetic modes of photonic waveguides in layered media.
Unlike finite difference or finite element methods which requires the volume
discretization and the imposition of artificial boundary conditions. Our
numerical scheme discretizes
the material interface only and treats the effect of layers efficiently via
the Sommerfeld representation, leading to a discrete linear system with very small
number of unknowns. The scheme is as well conditioned as the underlying physical
problem and thus capable of achieving high accuracy even for waveguides having
complex geometries and corners. Our benchmark computation on rectangular waveguides
imbedded in a cladding of finite thickness provides more than $10$ significant
digits for the propagation constants and shows quantitatively the relationship
between the modal confinement loss and the cladding thickness. The scheme, being
highly accurate and efficient for the mode calculation, provides a powerful design
and simulation tool for the photonics industry.

\bibliographystyle{abbrv}
\bibliography{journalnames,fmm,photonics,qbx}

\begin{thebibliography}{10}

\bibitem{srep}
S.~V. Boriskina, T.~M. Benson, P.~Sewell, and A.~I. Nosich.
\newblock Highly efficient full-vectorial integral equations method solution
  for the bound, leaky, and complex modes of dielectric waveguides.
\newblock {\em IEEE J. Selected Topics in Quantum Electron.}, 8:1225--1231,
  2002.

\bibitem{leslie1}
H.~Cheng, W.~Y. Crutchfield, M.~Doery, and L.~Greengard.
\newblock Fast, accurate integral equation methods for the analysis of photonic
  crystal fibers i: Theory.
\newblock {\em Optics Express}, 12(16):3791--3805, 2004.

\bibitem{fd3}
Y.~P. Chiou, Y.~C. Chiang, C.~H. Lai, C.~H. Du, and H.~C. Chang.
\newblock Finite difference modeling of dielectric waveguides with corners and
  slanted facets.
\newblock {\em J. Lightwave Technol.}, 27:2077--2086, 2009.

\bibitem{ferrando}
A.~Ferrando, E.~Silvestre, J.~Miret, P.~Andres, and M.~Andres.
\newblock Full vector analysis of a realistic photonic crystal fiber.
\newblock {\em Opt. Lett.}, 24:276--278, 1999.

\bibitem{kress}
R.~Kress.
\newblock {\em Linear Integral Equations}, volume~82 of {\em Applied
  Mathematical Sciences}.
\newblock Springer--Verlag, Berlin, 1989.

\bibitem{skie}
J.~Lai and S.~Jiang.
\newblock Second kind integral equation formulation for the mode calculation of
  optical waveguides.
\newblock {\em Appl. Comput. Harmon. Anal.}, accepted. {arxiv:1512.01117}.

\bibitem{lai}
J.~Lai, M.~Kobayashi, and L.~Greengard.
\newblock A fast solver for multi-particle scattering in a layered medium.
\newblock {\em Optics Express}, 14:302--307, 2014.

\bibitem{lu2}
W.~Lu and Y.~Y. Lu.
\newblock Waveguide mode solver based on {N}eumann-to-{D}irichlet operators and
  boundary integral equations.
\newblock {\em J. Comput. Phys.}, 231:1360--1371, 2012.

\bibitem{dmuller}
D.~E. M\"{u}ller.
\newblock A method for solving algebraic equations using an automatic computer.
\newblock {\em Math. Tables Aids Comput.}, 10:208--215, 1956.

\bibitem{fem6}
S.~S.~A. Obayya, B.~M.~A. Rahman, K.~T.~V. Grattan, and H.~A. El-Mikati.
\newblock Full vectorial finite-element-based imaginary distance beam
  propagation solution of complex modes in optical waveguides.
\newblock {\em J. Lightwave Technol.}, 20:1054--1060, 2002.

\bibitem{olver_nist_2010}
F.~W.~J. Olver, D.~W. Lozier, R.~F. Boisvert, and C.~W. Clark, editors.
\newblock {\em NIST Handbook of Mathematical Functions}.
\newblock Cambridge University Press, 2010.

\bibitem{pone}
E.~Pone, A.~Hassani, S.~Lacroix, A.~Kabashin, and M.~Skorobogatiy.
\newblock Boundary integral method for the challenging problems in bandgap
  guiding, plasmonics and sensing.
\newblock {\em Opt. Express}, 15:10231--10246, 2007.

\bibitem{fem5}
S.~Selleri, L.~V. L, A.~Cucinotta, and M.~Zoboli.
\newblock Complex {FEM} modal solver of optical waveguides with {PML} boundary
  conditions.
\newblock {\em Opt. Quant. Electron.}, 33:359--371, 2001.

\bibitem{fd2}
N.~Thomas, P.~Sewell, and T.~M. Benson.
\newblock A new full-vectorial higher order finite-difference scheme for the
  modal analysis of rectangular dielectric waveguides.
\newblock {\em J. Lightwave Technol.}, 25:2563--2570, 2007.

\bibitem{tpwhite1}
T.~P. White, B.~T. Kuhlmey, R.~C. McPhedran, D.~Maystre, G.~Renversez,
  C.~Martijn~de Sterke, and L.~C. Botten.
\newblock Multipole method for microstructured optical fibers. i. formulation.
\newblock {\em J. Opt. Soc. Am. B}, 19(10):2322--2330, 2002.

\end{thebibliography}

\end{document}